\documentclass[pra,a4,pacs]{revtex4}

\usepackage{amsfonts}
\usepackage{amsmath}
\usepackage{amssymb}
\usepackage{bm}
\usepackage[dvips]{color}

\usepackage{graphicx}
\usepackage{subfigure}

\newcommand{\hide}[1]{}

\begin{document}

%%%%%%%%%%%%%%%%%%%%%%%%%%%%%%%%%%%%%%%%%%%%%%%%%%%%%%%%%%%%%%%%%
%%%%%%%%%%%%%%%%%%%%%%%%%%%%%%%%%%%%%%%%%%%%%%%%%%%%%%%%%%%%%%%%%

\title{Molecules with an Induced Dipole Moment in a Stochastic
Electric Field}

\author{Y. B. Band}
\affiliation{Department of Chemistry, Department of Physics and
Department of Electro-Optics, and the Ilse Katz Center for Nano-Science,
Ben-Gurion University, Beer-Sheva 84105, Israel}
\author{Y. Ben-Shimol}
\affiliation{Department of Communication Systems Engineering,
Ben-Gurion University, Beer-Sheva 84105, Israel}

\date{\today}

\begin{abstract}
The mean-field dynamics of a molecule with an induced dipole moment
(e.g., a homonuclear diatomic molecule) in a deterministic and a
stochastic (fluctuating) electric field is solved to obtain the
decoherence properties of the system.  The average (over fluctuations)
electric dipole moment and average angular momentum as a function of
time for a Gaussian white noise electric field are determined via
perturbative and nonperturbative solutions in the fluctuating field.
In the perturbative solution, the components of the average electric
dipole moment and the average angular momentum along the deterministic
electric field direction do not decay to zero, despite fluctuations in
all three components of the electric field.  This is in contrast to
the decay of the average over fluctuations of a magnetic moment in a
stochastic magnetic field with a Gaussian white noise magnetic field
in all three components.  In the nonperturbative solution, the
component of the average electric dipole moment and the average
angular momentum in the deterministic electric field direction also
decay to zero.
\end{abstract}


\maketitle

\section{Introduction}  \label{Sec:Intro}

We consider the decoherence of a system, that has an induced electric
dipole moment, $d_i = \alpha_{ij} E_j$, where $\alpha_{ij}$ is the
polarizability tensor and $E_j$ is the $j$th component of an external
electric field, that is in contact with an environment (a bath).
Examples of such systems include homonuclear diatomic molecules, such
as H$_2$ \cite{Ishiguro_52} and N$_2$ \cite{Fleischer_07}, polyatomic
molecules with no permanent electric dipole moment (i.e., a molecule,
which, if fixed in space so that it cannot rotate, has a vanishing
electric dipole moment when no external electric field is present), or
a mezoscopic or macroscopic system, such as a colloidal particle
having no permanent dipole moment \cite{Scherer_04}.  The dynamics of
such systems that are in contact with an environment can be
represented by evolving the system in an effective electric field,
${\bf E}^{({\mathrm{eff}})} = {\bf E}_0 + {\bf E}_B(t)$, where ${\bf
E}_0$ is the deterministic electric field (which could be
time-dependent), and ${\bf E}_B(t)$ is the electric field which models
the influence of the environment (the bath $B$) on the dipole moment.
The field ${\bf E}_B(t)$ can be represented by a vector stochastic
process ${\boldsymbol \varepsilon}(t)$, where the nature of the
environment determines the type of stochastic process.  Averaging over
fluctuations corresponds to tracing over the environmental degrees of
freedom.  This yields a reduced nonunitary dynamics wherein the
averaged dipole moment and angular momentum decohere in time.  This
approach was recently used to treat decoherence of spin systems caused
by an environment \cite{STB_2013} and the decoherence of systems with
a permanent dipole moment \cite{Band_13}.  The physical properties of
the environment determine the statistical properties of ${\bf
E}_B(t)$, which in turn determine the type of stochastic process
${\boldsymbol \varepsilon}(t)$.  A prototype model for fluctuations is
Gaussian white noise, wherein the random process has vanishing
correlation time, but other types of noise are also commonly
encountered \cite{vanKampenBook, Kloeden, STB_2013}.

\section{Classical Dynamics}  \label{Sec:Cl_sol}

Consider a static electric field ${\bf E}$ in the direction of the
space-fixed $z$-axis and obtain the classical equations of motion of
the system.  The kinetic energy is $T=\tfrac{1}{2} I (\dot {\theta}^2
+ \sin^2 \theta \, \dot {\phi}^2)$ and the potential energy is $U = -
\frac{\alpha}{2} ({\bf n} \cdot {\bf E})^2 = - \frac{\alpha E^2}{2} \,
\cos^2 \theta$, where ${\bf n}$ is the unit vector in the direction of
the axis of the system (e.g., the unit vector along the axis of a
homonuclear diatomic molecule), hence the Lagrangian is,
\begin{equation} \label{Lagrangian}
    {\cal L}(\theta, \phi, \dot{\theta}, \dot {\phi}) = T - U =
    \tfrac{1}{2} I (\dot {\theta}^2+\sin^2 \theta \dot {\phi}^2) +
    \frac{\alpha E^2}{2} \, \cos^2 \theta .
\end{equation}
The Euler-Lagrange equations of motion are,
\begin{equation} \label{eq_phi}
    0 = \frac{\partial {\cal L}}{\partial \phi}-\frac{d}{dt}
    \frac{\partial {\cal L}}{\partial \dot{\phi}} = -I \frac{d}{dt}
    (\sin^2 \theta \, \dot{\phi}) \ \ \Rightarrow \ \ \dot{\phi}=
    \frac{\omega}{\sin^2 \theta}, \ \ \omega = \mbox{constant} ,
\end{equation} 
\begin{equation} \label{eq_theta}
    0 = \frac{\partial {\cal L}}{\partial \theta}-\frac{d}{dt}
    \frac{\partial {\cal L}}{\partial \dot{\theta}} \ \ \Rightarrow
    \quad \ddot{\theta} +\frac{\alpha E^2}{2I} \sin 2 \theta
    -\frac{\omega^2}{2} \frac{\sin 2 \theta}{\sin^2 \theta} = 0 .
\end{equation} 
The dynamics are relatively simple since the $z$-component of the
angular momentum, $L_z = \frac{\partial {\cal L}}{\partial \dot{\phi}}
= I \sin^2 \theta \, \dot{\phi} \equiv I \omega$, is conserved because
there is no component of torque, ${\boldsymbol \tau} = {\bf d} \times
{\bf E} = \frac{\alpha}{2} (\hat{{\bf n}} \cdot {\bf E}) (\hat{{\bf
n}} \times {\bf E})$, along $z$.  The second constant of the
motion is the total energy ${\cal E}$,
\begin{equation} \label{d}
    {\cal E} = T + U = \tfrac{1}{2} I (\dot {\theta}^2 +
    \frac{\omega^2}{ \sin^2 \theta} ) - \frac{\alpha E^2}{2} \, \cos^2
    \theta ~.
\end{equation}

\begin{figure} % [hbt]
\centering
\centering\subfigure[]{\includegraphics[width=0.45\textwidth]
{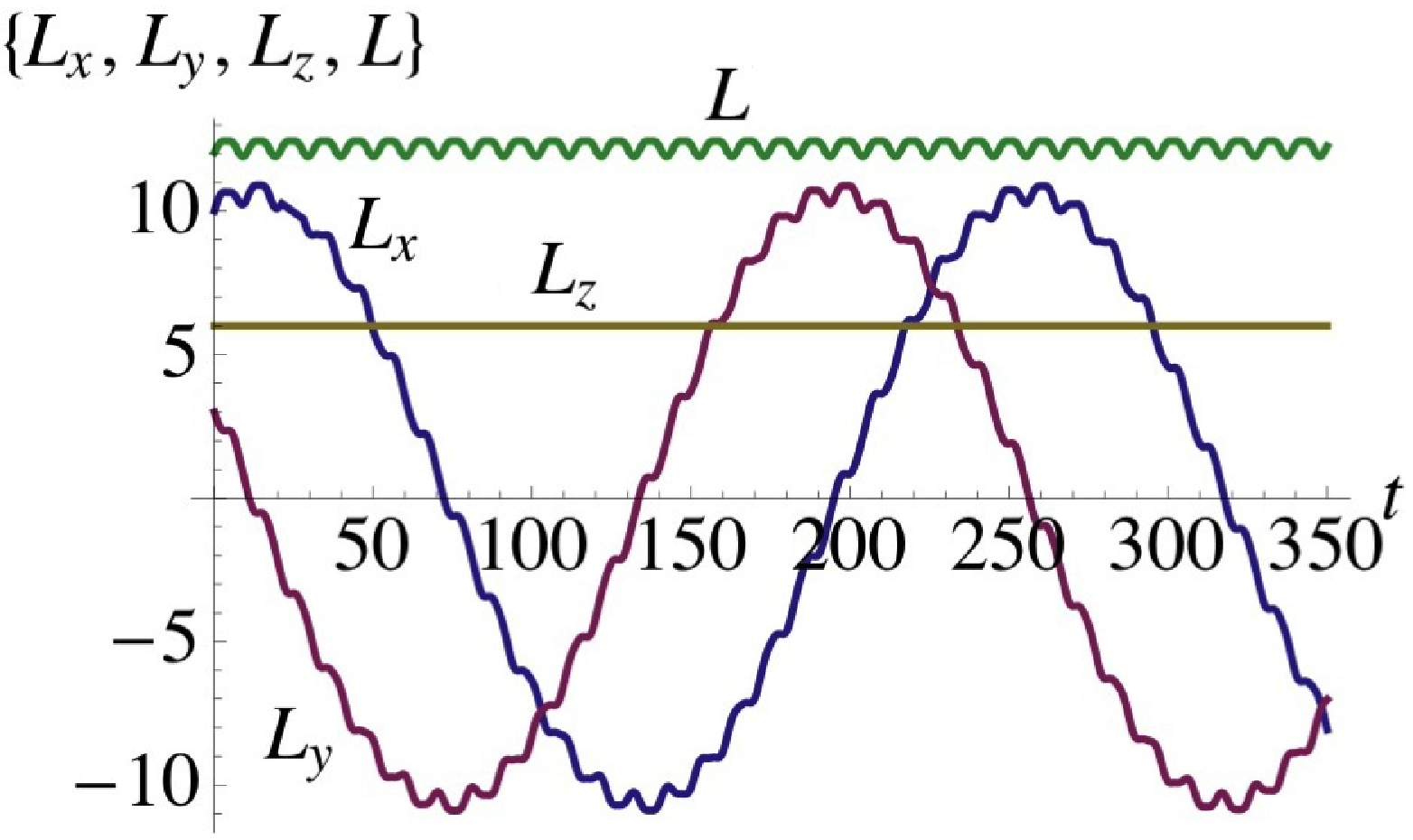}}
\centering\subfigure[]{\includegraphics[width=0.35\textwidth]
{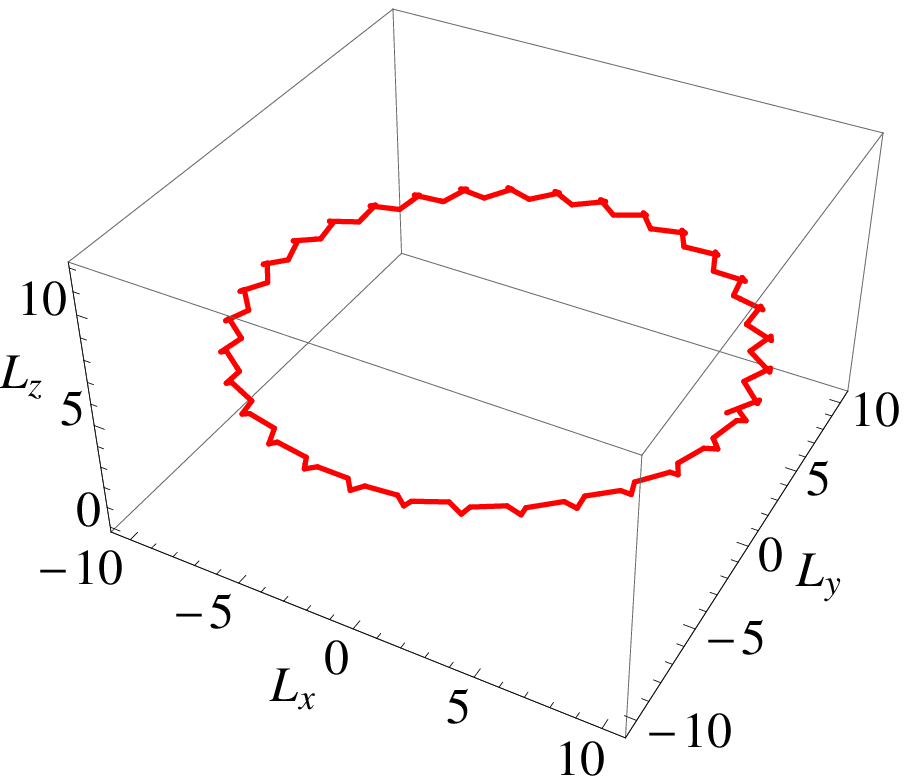}} 
\caption{(Color online) $L_x(t), L_y(t), L_z(t)$ and $L(t) =
\sqrt{L^2_x(t) + L^2_y(t) + L^2_z(t)}$ (green curve) versus for ${\bf
n}(0) = (\sin(\pi/4) \cos(\pi/4),\sin(\pi/4) \sin(\pi/4),\cos(\pi/4))$
and $(L_x(0), L_y(0), L_z(0)) = (10, 3, 6)$.}
\label{Fig_induced_dipole}
\end{figure}

\begin{figure} % [hbt]
\centering
\includegraphics[width=0.5\textwidth]{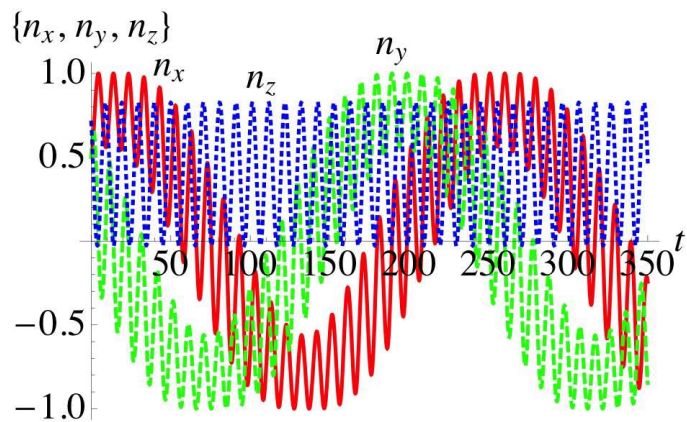}
\caption{(Color online) $n_x(t), n_y(t)$, and $n_z(t)$ versus time for
$n_x(0) = n_y(0) = 1/2$, $n_z(0) = 1/\sqrt{2}$ (initial conditions for
${\bf L}(t)$ given in Fig.~\ref{Fig_induced_dipole} caption).}
\label{Fig_Induced_dipole_n}
\end{figure}

\begin{figure} % [hbt]
\centering
\includegraphics[width=0.5\textwidth]{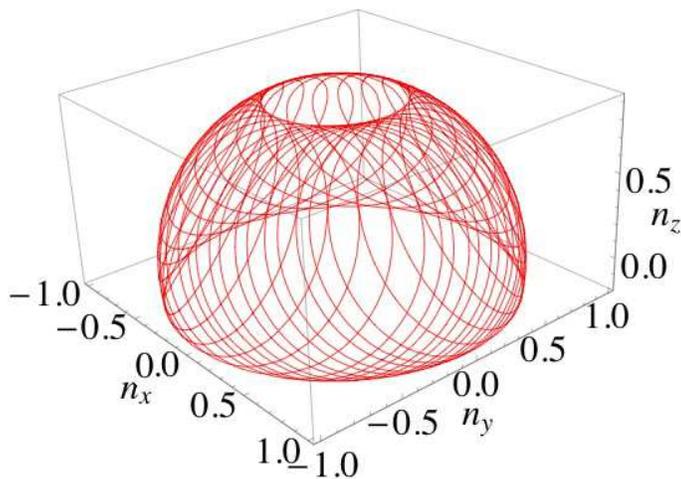}
\caption{(Color online) Parametric plot of ${\bf n}(t)$ versus time.}
\label{Fig_Induced_dipole_n_parametric}
\end{figure}

Another way of expressing the classical equations of motion is in
terms of the angular momentum ${\bf L}$ and the unit vector ${\bf n}$,
\begin{equation} \label{dot_n}
    \dot{{\bf n}} = - \frac{1}{I} {\bf L} \times {\bf n} =
    - {\boldsymbol \omega} \times {\bf n} .
\end{equation}
\begin{equation} \label{dot_L}
    \dot{{\bf L}} = - \alpha ({\bf n} \cdot
    {\bf E}) ({\bf E} \times {\bf n}) .
\end{equation}
Figure~\ref{Fig_induced_dipole} shows $L_x(t), L_y(t), L_z(t)$, and
$L(t) = \sqrt{L^2_x(t) + L^2_y(t) + L^2_z(t)}$ versus time, and
Figs.~\ref{Fig_Induced_dipole_n} and
\ref{Fig_Induced_dipole_n_parametric} show $n_x(t), n_y(t)$, and
$n_z(t)$ versus time.  The dimensionless parameters used in these
calculations are $\alpha = 1$, $I = 20$, and $(E_x, E_y, E_z) =
(0,0,1)$, and the initial conditions are taken as, $(L_x(0), L_y(0),
L_z(0)) = (10, 3, 6)$, and ${\bf n}(0) = (\sin(\pi/4)
\cos(\pi/4),\sin(\pi/4) \sin(\pi/4),\cos(\pi/4)) = (1/2, 1/2,
1/\sqrt{2})$.  The dynamics is {\em almost} periodic with period of
about 250 (dimensionless units).

\section{Quantum Treatment}

The Hamiltonian for the system is given by $H = \frac{ {\hat L}^2}{2
I} - \hat{{\bf d}} \cdot {\bf E}$; taking the induced dipole moment
operator to have a component only along the system axis, $\hat{{\bf
d}} = (\alpha/2) (\hat{{\bf n}}\cdot {\bf E}) \hat{{\bf n}}$, where
$\hat{{\bf n}}$ is a vector operator of unit length in the direction
of the system axis, we obtain \cite{sym_top},
\begin{equation}  \label{Ham_sym_top_E_field'}
    H = \frac{{\hat L}^2}{2 I} - \frac{\alpha}{2} (\hat{{\bf n}} \cdot
    {\bf E})^2 .
\end{equation}
The Heisenberg equation of motion for $\hat{{\bf n}}$ and $\hat{{\bf
L}}$, $\dot{\hat{{\bf n}}} = \frac{i}{\hbar} [H, \hat{{\bf n}}]$ and
$\dot{\hat{{\bf n}}} = \frac{i}{\hbar} [H, \hat{{\bf n}}]$, determine
the dynamics.  For $\dot {\hat{{\bf n}}}$ we find,
\begin{equation} \label{dot_n'}
    \dot{\hat{{\bf n}}} = \frac{i}{2 \hbar I}[{\hat L}^2, \hat{{\bf
    n}}] - \frac{i \, \alpha}{2 \hbar} [(\hat{{\bf n}} \cdot {\bf
    E})^2, \hat{{\bf n}}] .
\end{equation}
Using the fact that $[\hat{L}_i, \hat{n}_j] = i \hbar \epsilon_{ijk}
\hat{n}_k$, we obtain, $[{\hat L}^2, \hat{{\bf n}}] = 2 i \hbar [
\hat{{\bf L}} \times \hat{{\bf n}} + i \hbar \, \hat{{\bf n}} ]$.
Because $[\hat{n}_i, \hat{n}_j] = 0$ for all $i$ and $j$, the second
term on the RHS of Eq.~(\ref{dot_n'}) vanishes, and we find,
\begin{equation} \label{dot_n_final'}
    \dot{\hat{{\bf n}}} = - \frac{1}{I} [ \hat{{\bf L}} \times
    \hat{{\bf n}} + i \hbar \, \hat{{\bf n}} ] .
\end{equation}
The torque on the molecule due to the presence of the external field,
$\dot{\hat{{\bf L}}} = \frac{i}{\hbar} [H, \hat{{\bf L}}]$, is given
by
\begin{equation} \label{dot_L'}
    \dot{\hat{{\bf L}}} = - \alpha (\hat{{\bf n}} \cdot
    {\bf E}) ({\bf E} \times \hat{{\bf n}}) .
\end{equation}
Since the angular momentum is not conserved, the solution of the
Heisenberg equations of motion would require a basis set calculation
including many angular momentum states; doing so with a stochastic
electric field (see Sec.~\ref{Sec:Stochastic}) would be very 
tedious.  Therefore, we develop a mean-field approach.

\subsection{Mean-Field Dynamics}  \label{SubSec:MFD}

If the initial angular momentum of the molecule is large compared to
$\hbar$, a semiclassical treatment can be a good approximation.
Setting $\hbar = 0$ in Eq.~(\ref{dot_n_final'}) allows a semiclassical
solution for the expectation values $\langle {\hat{{\bf n}}}(t)
\rangle$ and $\langle {\hat{{\bf L}}}(t) \rangle$.  The semiclassical
equations are equivalent to the classical solution presented in
Sec.~\ref{Sec:Cl_sol}, and are valid for arbitrary direction of ${\bf
E}$.  The mean-field theory treatment takes the expectation values of
Eqs.~(\ref{dot_n_final'}) and (\ref{dot_L'}), replacing the
expectation value of the product $\hat{{\bf L}} \times \hat{{\bf n}}$
by the product of the expectation values \cite{Zobay_00, Liu_02,
Tikhonenkov_06, Band_07} and taking the limit as $\hbar \to 0$ on the
RHS of (\ref{dot_n_final'}):
\begin{equation} \label{dot_n_final}
     \langle \dot{\hat{{\bf n}}} \rangle = - \frac{1}{I}  \langle 
     {\hat{{\bf L}}} \rangle \times
     \langle {\hat{{\bf n}}} \rangle,
\end{equation}
\begin{equation} \label{dot_L_final}
    \langle \dot{\hat{{\bf L}}} \rangle = - \alpha ( \langle
    {\hat{{\bf n}}} \rangle \cdot {\bf E}) ({\bf E} \times \langle
    {\hat{{\bf n}}} \rangle) .
\end{equation}
The nonlinear equations of motion, (\ref{dot_n_final}) and
(\ref{dot_L_final}) [which are the same as Eqs.~(\ref{dot_n}) and
(\ref{dot_L})] must be solved simultaneously.

\section{Stochastic Dynamics} \label{Sec:Stochastic}

We now consider the dynamics in the presence of a stochastic electric
field, so the total electric field is taken to be the sum of a
deterministic field and a stochastic field, ${\bf E} = {\bf E}_0 +
{\boldsymbol \varepsilon}(t)$, where ${\boldsymbol \varepsilon}(t)$ is
a stochastic process.  We solve for the dynamics in two ways.  First
we treat the stochastic field perturbatively, by dropping the term in
the dynamical equations of motion that is quadratic in ${\boldsymbol
\varepsilon}(t)$ and by taking the linear term in ${\boldsymbol
\varepsilon}(t)$ to be Gaussian white noise.  Then we treat the full 
(nonperturbative) dynamics, taking ${\boldsymbol \varepsilon}(t)$ to 
be an Ornstein-Uhlenbeck process.

In what follows, we denote the quantum averages of the unit vector
along the axis of the molecule and the angular momentum by ${\bf n}(t)
\equiv \langle {\hat{{\bf n}}}(t) \rangle$ and ${\bf L}(t) \equiv
\langle {\hat{{\bf L}}}(t) \rangle$.  The average of these quantities 
over the stochasticity can be denoted by $\overline{{\bf n}(t)}$ and 
$\overline{{\bf L}(t)}$ respectively.

\subsection{Perturbation Theory in ${\boldsymbol \varepsilon}(t)$}
\label{SubSec:Perturbation}

In Eq.~(\ref{dot_L_final}), we substitute ${\bf E} = {\bf E}_0 +
{\boldsymbol \varepsilon}(t)$, and expand, keeping only the linear
term in ${\boldsymbol \varepsilon}(t)$ and dropping the quadratic
term.  The resulting equation is of the form of a stochastic
differential equation.  We need only specify the details of the
stochastic electric field ${\boldsymbol \varepsilon}(t)$.  We take
$\varepsilon_x(t)$, $\varepsilon_y(t)$ and $\varepsilon_z(t)$ to be
stochastic processes with zero mean and delta function correlation
function $\kappa(t-t')$,
\begin{equation}  \label{white_noise_1}
    \overline{\varepsilon_i(t)} = 0 ,
\end{equation}
\begin{equation}  \label{white_noise_2}
    \overline{\varepsilon_i(t) \varepsilon_j(t')} = \kappa(t-t') \, 
    \delta_{ij} = \varepsilon_0^2 \, \delta (t-t') \, \delta_{ij} ,
\end{equation}
for $i, j = x, y$ and $z$, i.e., we consider a vector Wiener process.
Equations (\ref{dot_n_final}) and (\ref{dot_L_final}) form a 
system of differential equations, which can be written in the 
standard stochastic differential equation form \cite{vanKampenBook},
\begin{eqnarray}
    d{\bf n}(t) &=& - \frac{1}{I} {\bf L}(t) \times {\bf n}(t) \, dt ,
    \label{dot_dn_final}  \\
    d{\bf L}(t) &=& - \alpha \, [({\bf n}(t) \cdot {\bf E}_0 ) \,
    ({\bf E}_0 \times {\bf n}(t)) \, dt + ({\bf n}(t) \cdot {\bf E}_0)
    \, ( d{\boldsymbol \varepsilon}(t) \times {\bf n}(t)) + ({\bf n}(t)
    \cdot d{\boldsymbol \varepsilon}(t)) \, ({\bf E}_0 \times {\bf
    n}(t))] , \label{dot_dL}
\end{eqnarray}

\begin{figure} % [hbt]
\centering
\centering\subfigure[]{\includegraphics[width=0.3\textwidth]
{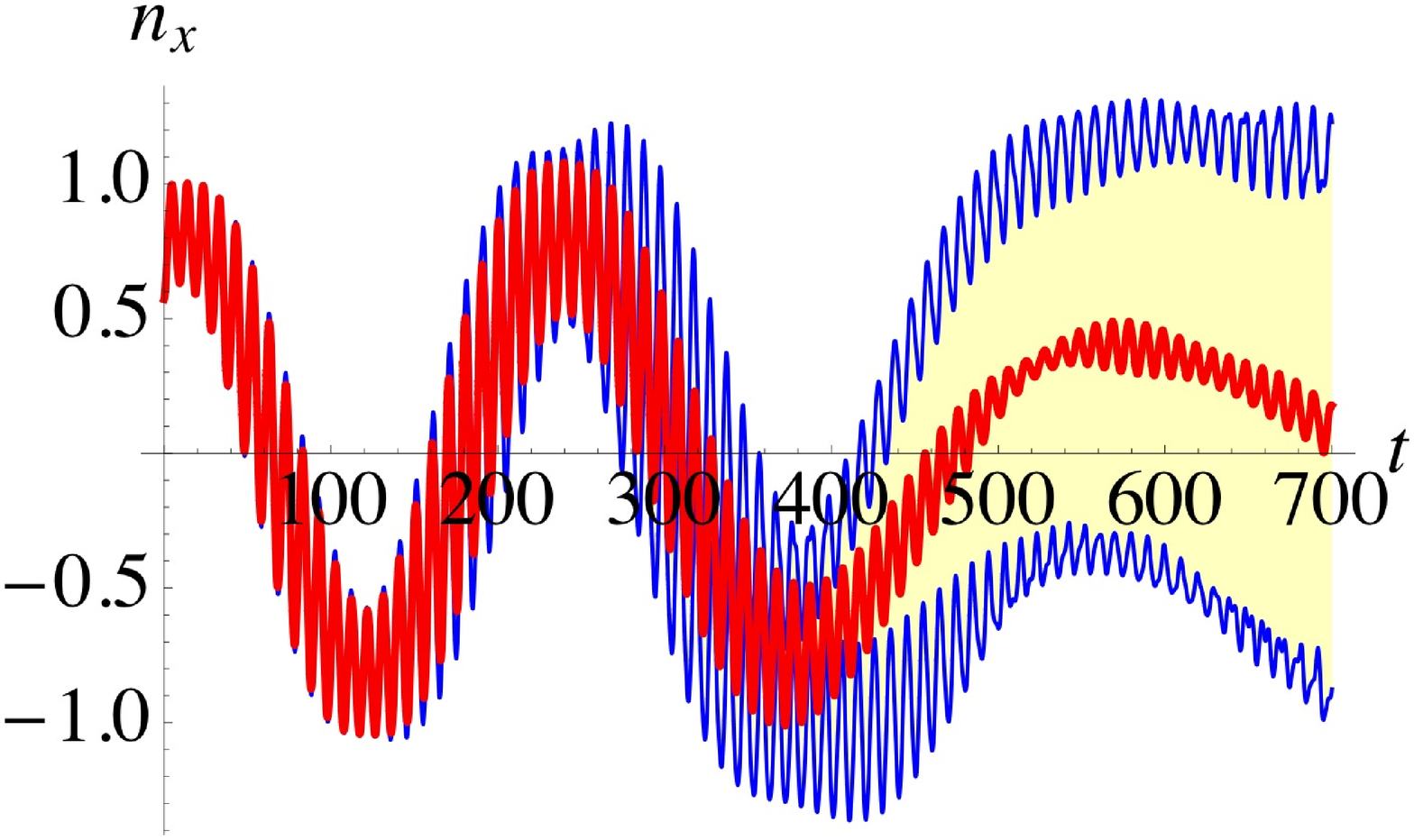}}
\centering\subfigure[]{\includegraphics[width=0.3\textwidth]
{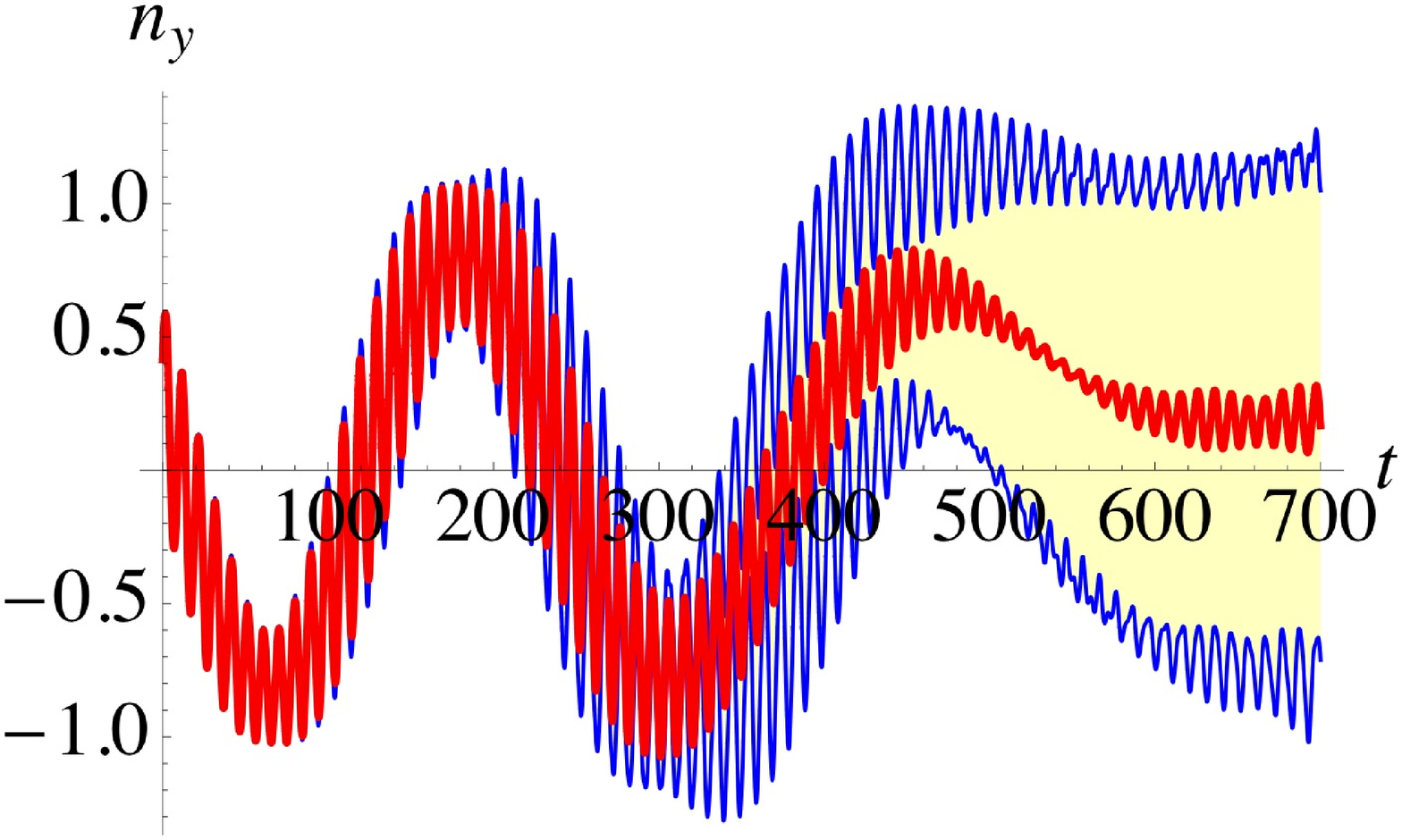}} 
\centering\subfigure[]{\includegraphics[width=0.3\textwidth]
{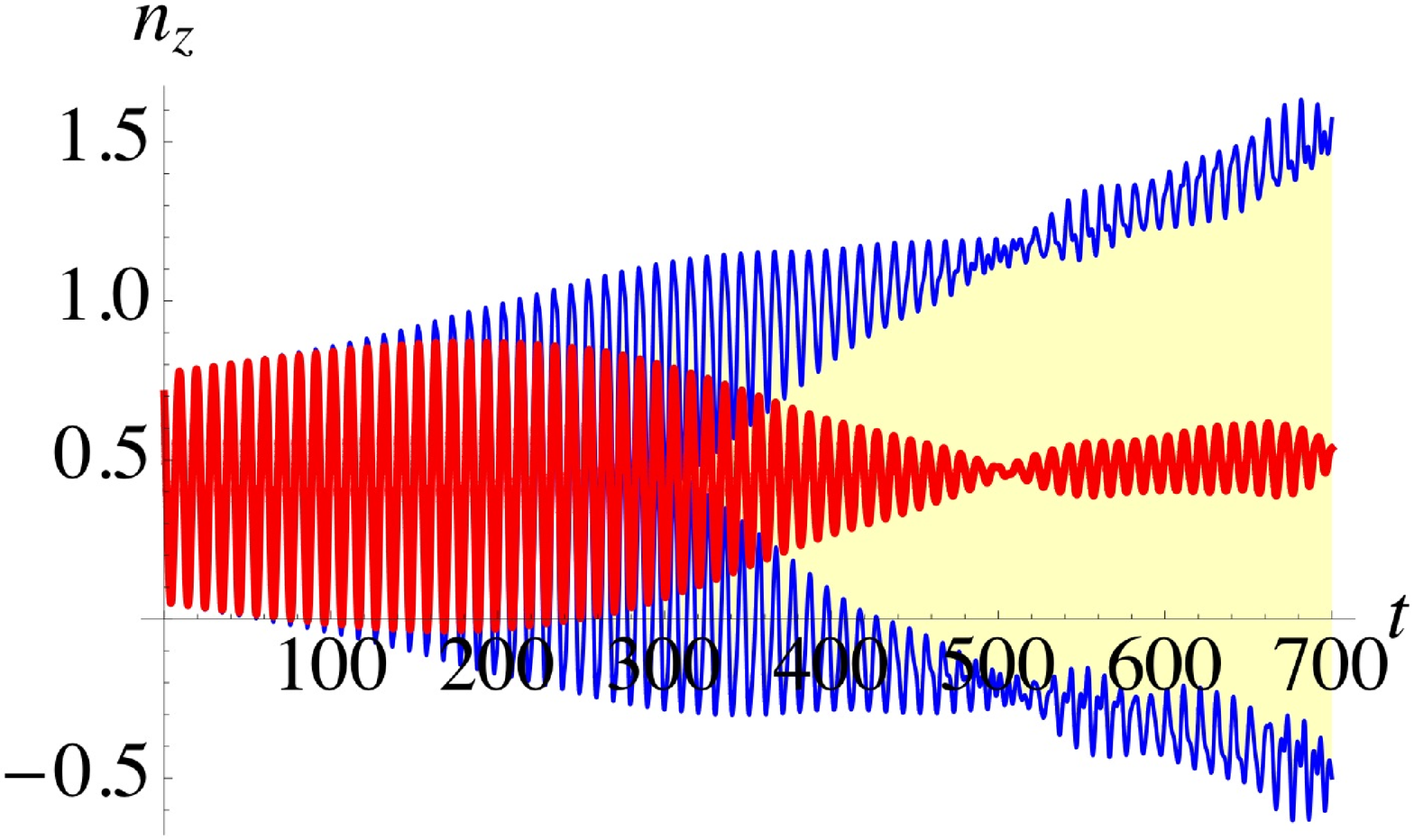}} 
\caption{(Color online) Average and standard deviation of $n_x(t),
n_y(t), n_z(t)$ versus time obtained for stochastic dynamics using
Eqs.~(\ref{dot_dn_final}) and (\ref{dot_dL}) with $\varepsilon_x(t)$,
$\varepsilon_y(t)$ and $\varepsilon_z(t)$ fields taken as Gaussian
white noise and $\sigma = 0.02$.}
\label{Fig_vareps_z_n_m_v_0.02}
\end{figure}

\begin{figure} % [hbt]
\centering
\centering\subfigure[]{\includegraphics[width=0.3\textwidth]
{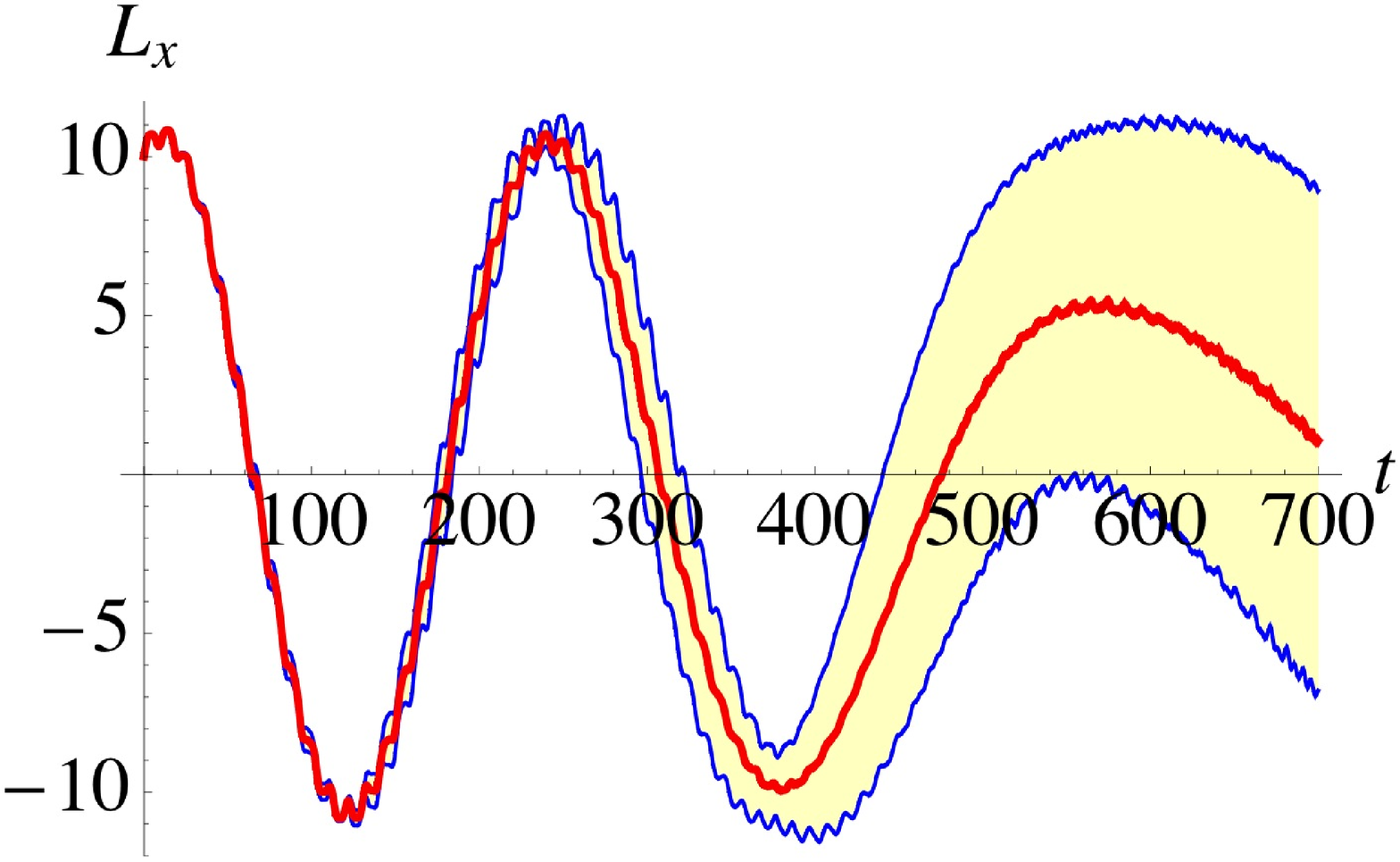}}
\centering\subfigure[]{\includegraphics[width=0.3\textwidth]
{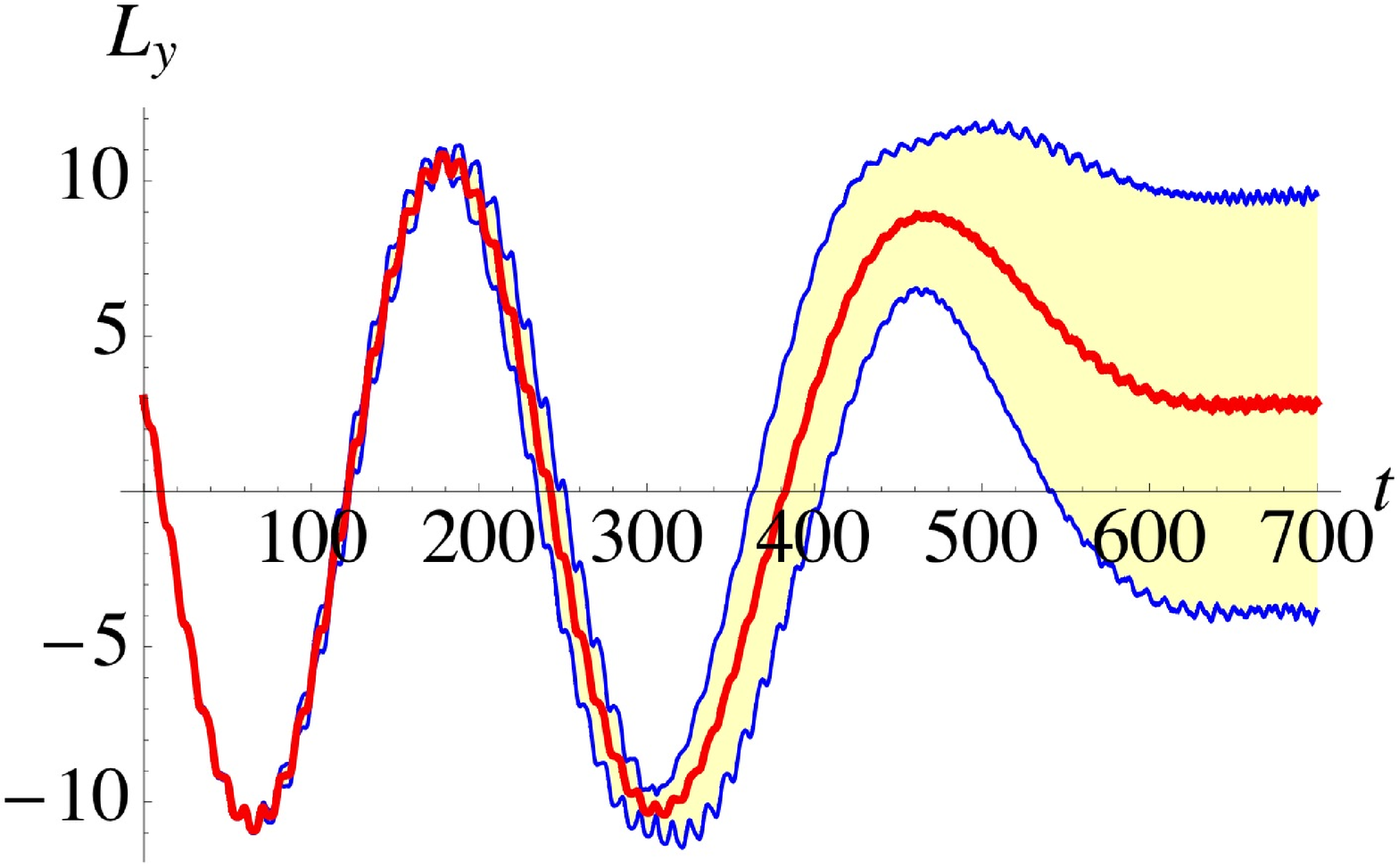}} 
\centering\subfigure[]{\includegraphics[width=0.3\textwidth]
{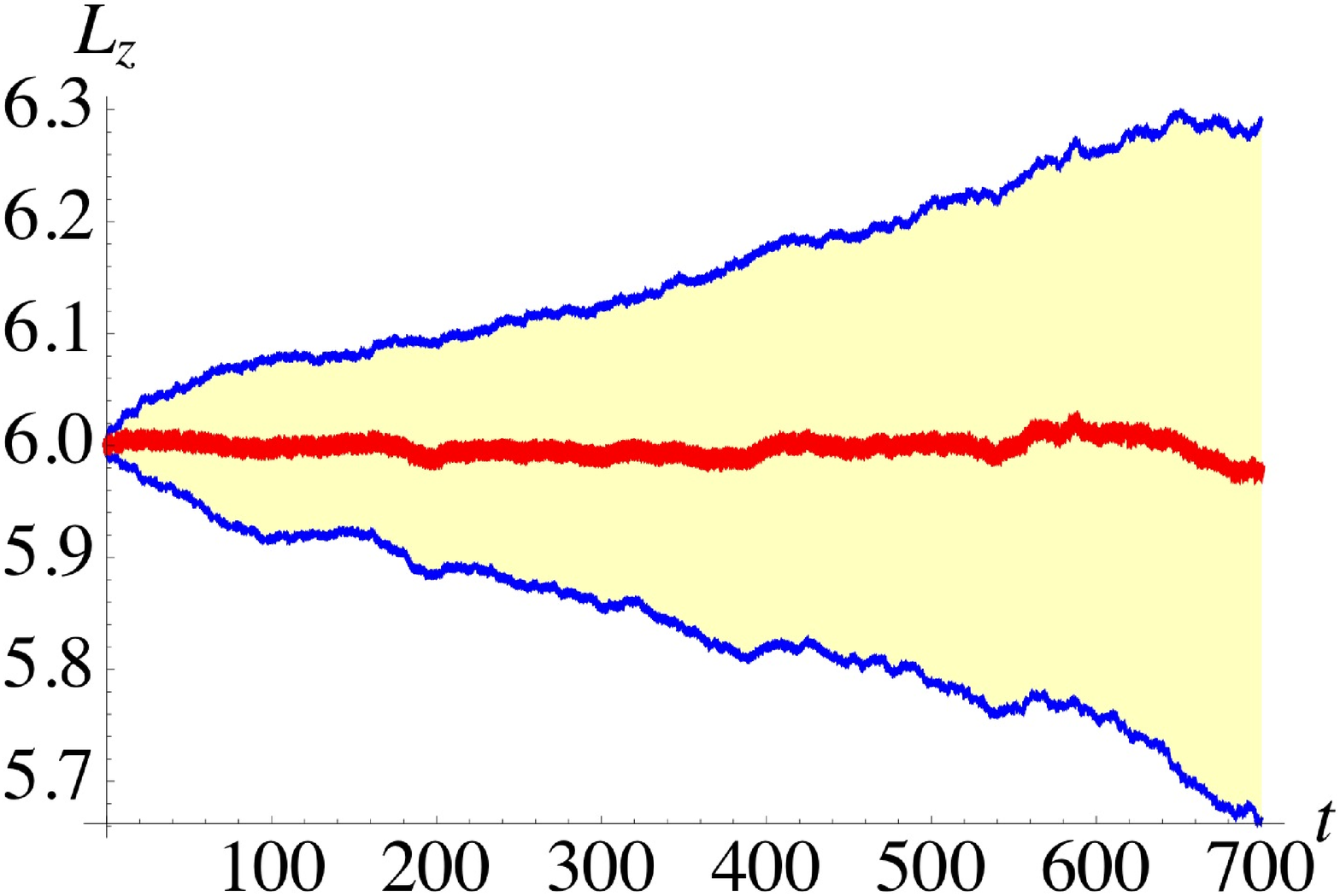}} 
\caption{(Color online) Average and standard deviation of the angular
momentum vector $(L_x(t), L_y(t), L_z(t))$ versus time obtained for
stochastic dynamics using Eqs.~(\ref{dot_dn_final}) and (\ref{dot_dL})
with $\varepsilon_x(t)$, $\varepsilon_y(t)$ and $\varepsilon_z(t)$
fields taken as Gaussian white noise and $\sigma = 0.02$.}
\label{Fig_vareps_z_L_m_v_0.02}
\end{figure}

In the numerical calculations we took ${\bf E}_0 = (0,0,1)$ (in
dimensionless units) and the initial conditions ${\bf n}(0)$ and ${\bf
L}(0)$ as in Figs.~\ref{Fig_induced_dipole},
\ref{Fig_Induced_dipole_n} and \ref{Fig_Induced_dipole_n_parametric}.
Figures \ref{Fig_vareps_z_n_m_v_0.02} and
\ref{Fig_vareps_z_L_m_v_0.02} show the vectors ${\bf n}(t) = (n_x(t),
n_y(t), n_z(t))$ and ${\bf L}(t) = (L_x(t), L_y(t), L_z(t))$
calculated with the stochastic fields, $\varepsilon_x(t)$,
$\varepsilon_y(t)$ and $\varepsilon_z(t)$ taken as Gaussian white
noise with a small stochastic field strength (volatility),
$\varepsilon_0 = 0.02$.  The central curves (in red) give the averages
$\overline{{\bf n}(t)}$ and $\overline{{\bf L}(t)}$ in
Figs.~\ref{Fig_vareps_z_n_m_v_0.02} and \ref{Fig_vareps_z_L_m_v_0.02},
and the mean values plus and minus the standard deviations are shown
as curves (in blue), with the region between the plus and minus
standard deviations shaded (in yellow).  Since the initial conditions
for the stochastic field is taken to be ${\boldsymbol \varepsilon}(0)
= {\bf 0}$, and $\varepsilon_0$ is small, the dynamical variables
${\bf n}(t)$ and ${\bf L}(t)$ start off very much like the variables
calculated without stochasticity, but by a time of about 250
(dimensionless time units), decoherence is evident.  The decoherence
becomes significant for times larger than about 400.  The variables
$n_x(t), n_y(t)$ and $L_x(t), L_y(t)$ decay to zero at large times,
but $n_z(t)$ and $L_z(t)$ `hang up' at finite values.  It is clear
from Eq.~(\ref{dot_dL}) that $d\langle L_z(t) \rangle/dt = 0$, because
$d\langle {\boldsymbol \varepsilon}(t) \rangle/dt = 0$ and $({\bf E}_0
\times {\bf n}(t))_z = 0$.

\begin{figure} % [hbt]
\centering
\centering\subfigure[]{\includegraphics[width=0.3\textwidth]
{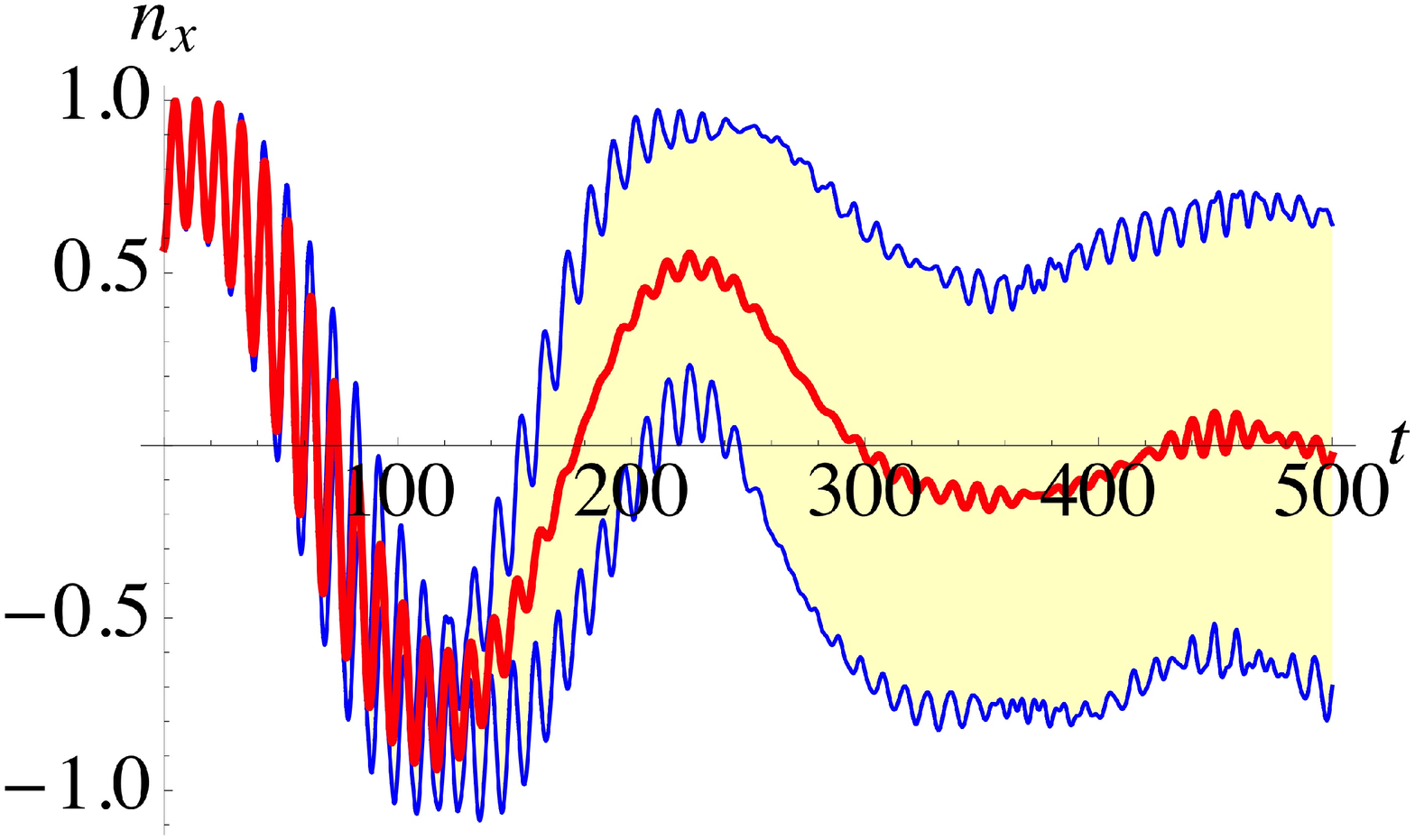}}
\centering\subfigure[]{\includegraphics[width=0.3\textwidth]
{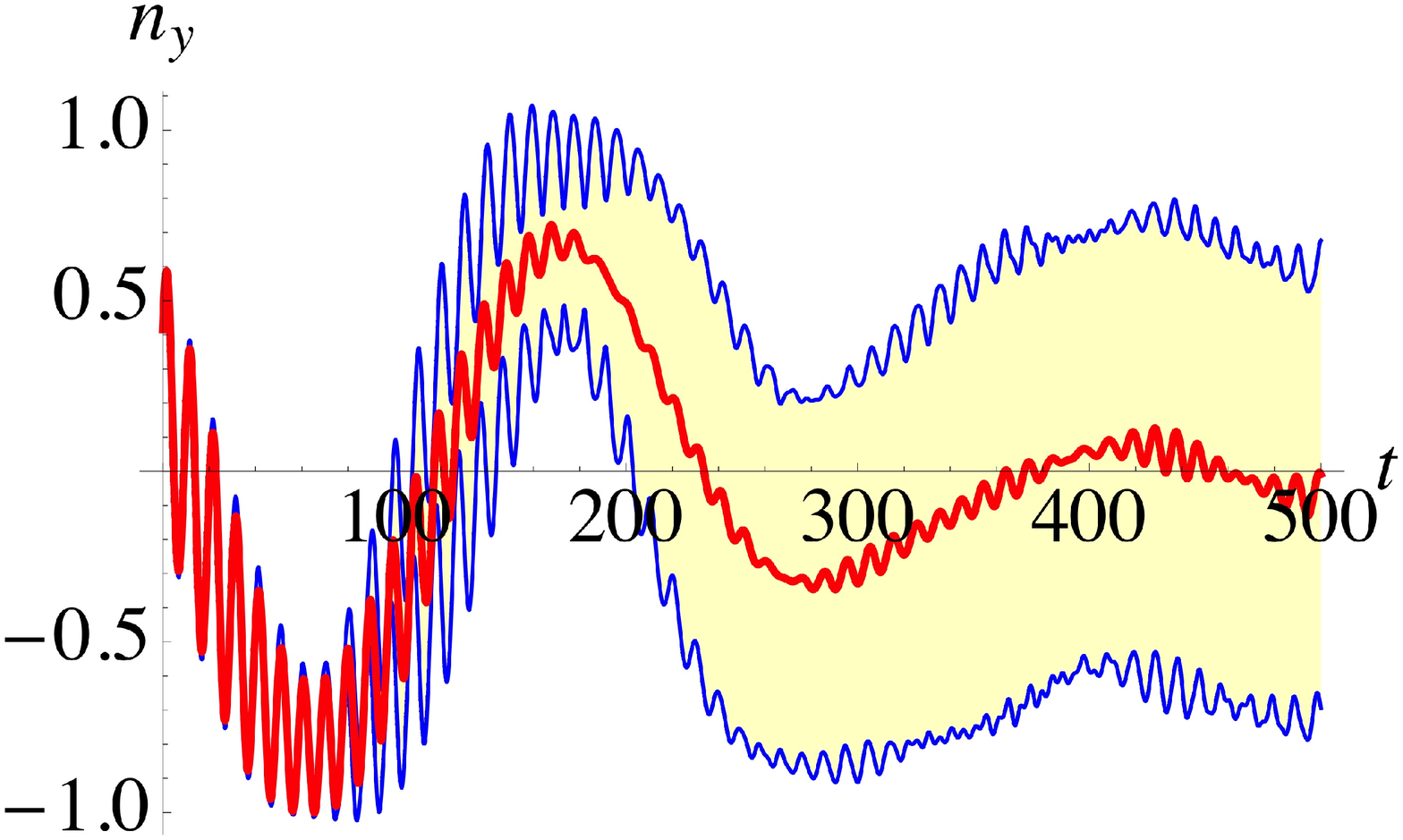}} 
\centering\subfigure[]{\includegraphics[width=0.3\textwidth]
{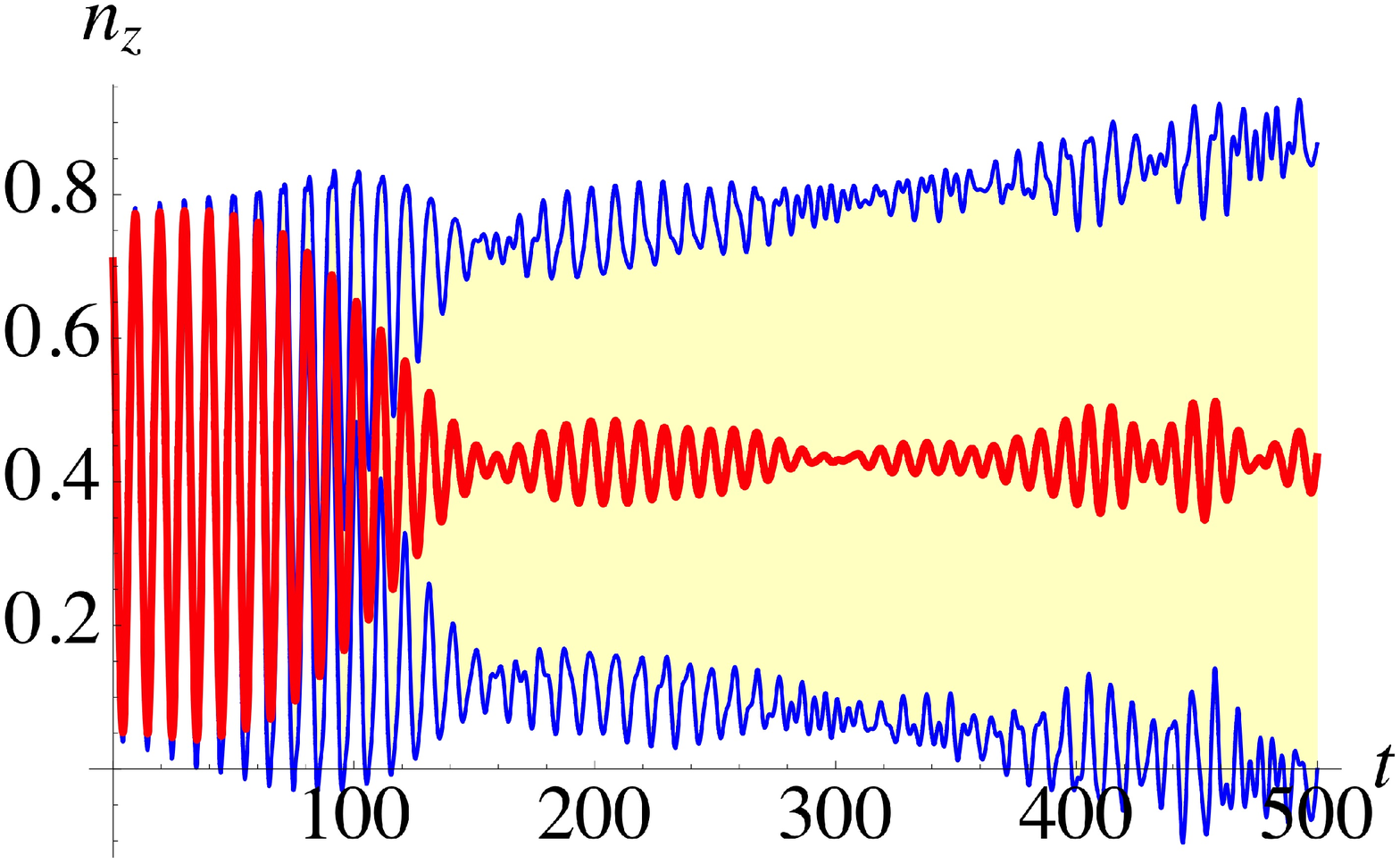}} 
\caption{(Color online) Average and standard deviation of $n_x(t),
n_y(t), n_z(t)$ versus time obtained for stochastic dynamics using
Eqs.~(\ref{dot_dn_final}) and (\ref{dot_dL}) with $\varepsilon_x(t)$,
$\varepsilon_y(t)$ and $\varepsilon_z(t)$ fields taken as Gaussian
white noise and $\sigma = 0.1$.}
\label{Fig_vareps_z_n_m_v}
\end{figure}

\begin{figure} % [hbt]
\centering
\centering\subfigure[]{\includegraphics[width=0.3\textwidth]
{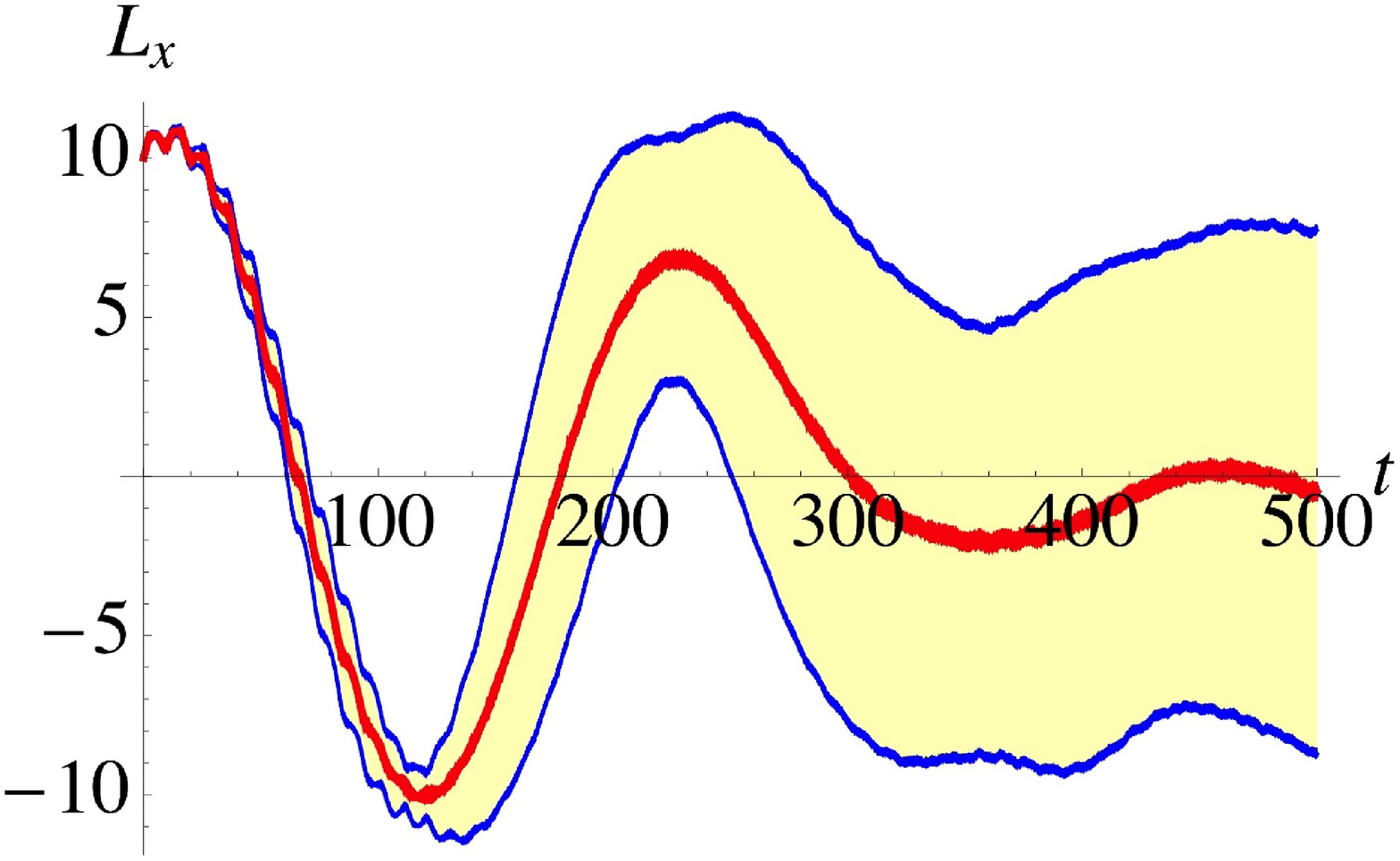}}
\centering\subfigure[]{\includegraphics[width=0.3\textwidth]
{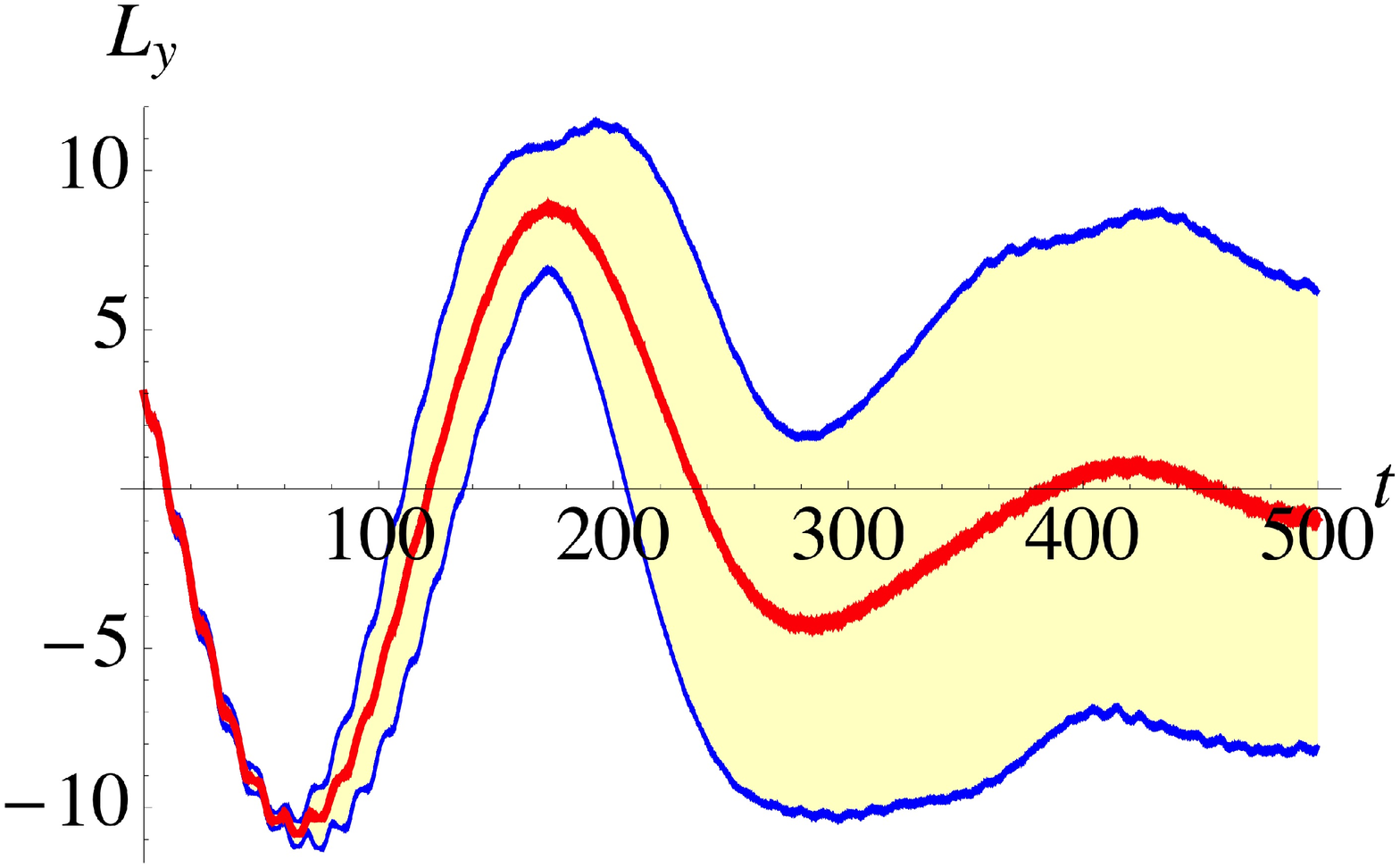}} 
\centering\subfigure[]{\includegraphics[width=0.3\textwidth]
{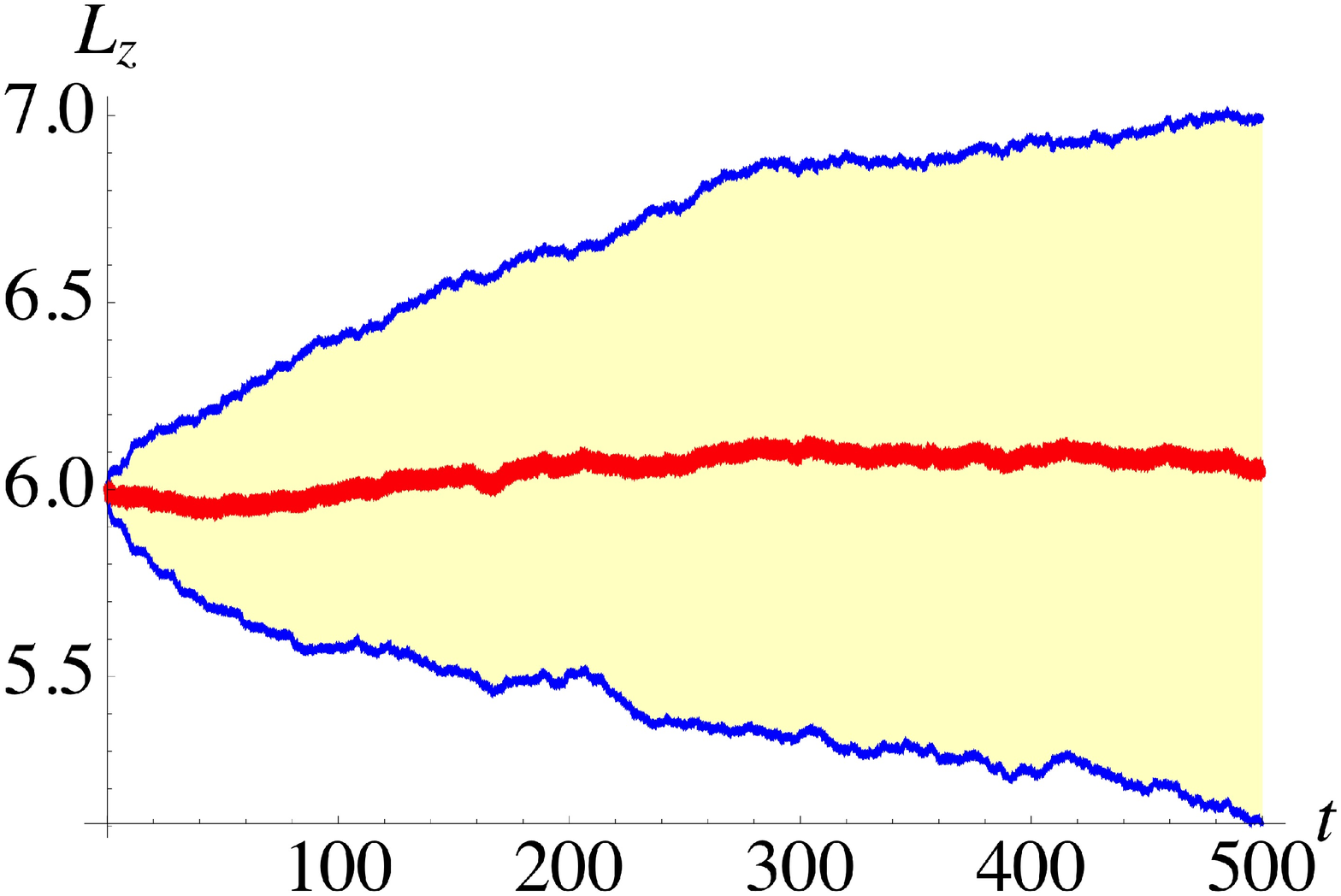}} 
\caption{(Color online) Average and standard deviation of the angular
momentum vector $(L_x(t), L_y(t), L_z(t))$ versus time obtained for
stochastic dynamics using Eqs.~(\ref{dot_dn_final}) and (\ref{dot_dL})
with $\varepsilon_x(t)$, $\varepsilon_y(t)$ and $\varepsilon_z(t)$
fields taken as Gaussian white noise and $\sigma = 0.02$.}
\label{Fig_vareps_z_L_m_v}
\end{figure}

Figures \ref{Fig_vareps_z_n_m_v} and \ref{Fig_vareps_z_L_m_v} are
similar to Figs.~\ref{Fig_vareps_z_n_m_v_0.02} and
\ref{Fig_vareps_z_L_m_v_0.02} and show the average and standard
deviation of the vectors ${\bf n}(t) = (n_x(t), n_y(t), n_z(t))$ and
${\bf L}(t) = (L_x(t), L_y(t), L_z(t))$ calculated with a larger value
of volatility, $\varepsilon_0 = 0.1$.  Now, decoherence sets in at
earlier times, becoming significant for times larger than around 150.
Again, $\overline{n_z(t)}$ and $\overline{L_z(t)}$ `hang up' at finite
values.  Since we expect perturbation theory to begin to break down at
larger values of volatility, we now carry out nonperturbative
calculations.

\subsection{Nonlinear in ${\boldsymbol \varepsilon}(t)$ Calculation}
\label{SubSec:Perturbation}

In order to treat the equations of motion that are nonlinear in the
variable ${\boldsymbol \varepsilon}$, we recast them in the form
[recall that we are using the notation ${\bf n}(t) \equiv \langle
{\hat{{\bf n}}}(t) \rangle$ and ${\bf L}(t) \equiv \langle {\hat{{\bf
L}}}(t) \rangle$],
\begin{eqnarray}
    d{\bf n}(t) &=& - \frac{1}{I} {\bf L}(t) \times {\bf n}(t) \, dt ,
    \label{dot_dn_final_nl}  \\
    d{\bf L}(t) &=& - \alpha \, {\bf n}(t) \cdot [{\bf E}_0 + {\boldsymbol
	\varepsilon}(t)] \, [{\bf E}_0 + {\boldsymbol \varepsilon}(t)]
	\times {\bf n}(t) \, dt , \label{dot_dL_nl} \\
    d {\boldsymbol \varepsilon}(t) &=& -\vartheta {\boldsymbol
	\varepsilon}(t) \, dt + \sigma d {\bf W}(t).  \label{d_varepsilon}
\end{eqnarray}
Here $\vartheta$ is the mean reversion rate of the Ornstein-Uhlenbeck
process ${\boldsymbol \varepsilon}$ with long term mean of equal zero,
${\bf W}(t)$ is the standard Wiener process with zero mean and
volatility one, and $\sigma$ is the volatility of the
Ornstein-Uhlenbeck process.  With nonvanishing $\vartheta$, the
functional dependence on time of the variance and correlation function
of the process ${\boldsymbol \varepsilon}$ is very different from a
Wiener process, but with $\vartheta = 0$, the process ${\boldsymbol
\varepsilon}$ {\em is} a Wiener process.  In the numerical
calculations presented here, we take $\vartheta = 0$ and initial
conditions $\varepsilon_i(0) = 0$ for $i = 1, 2, 3$.

\begin{figure} % [hbt]
\centering
\centering\subfigure[]{\includegraphics[width=0.3\textwidth]
{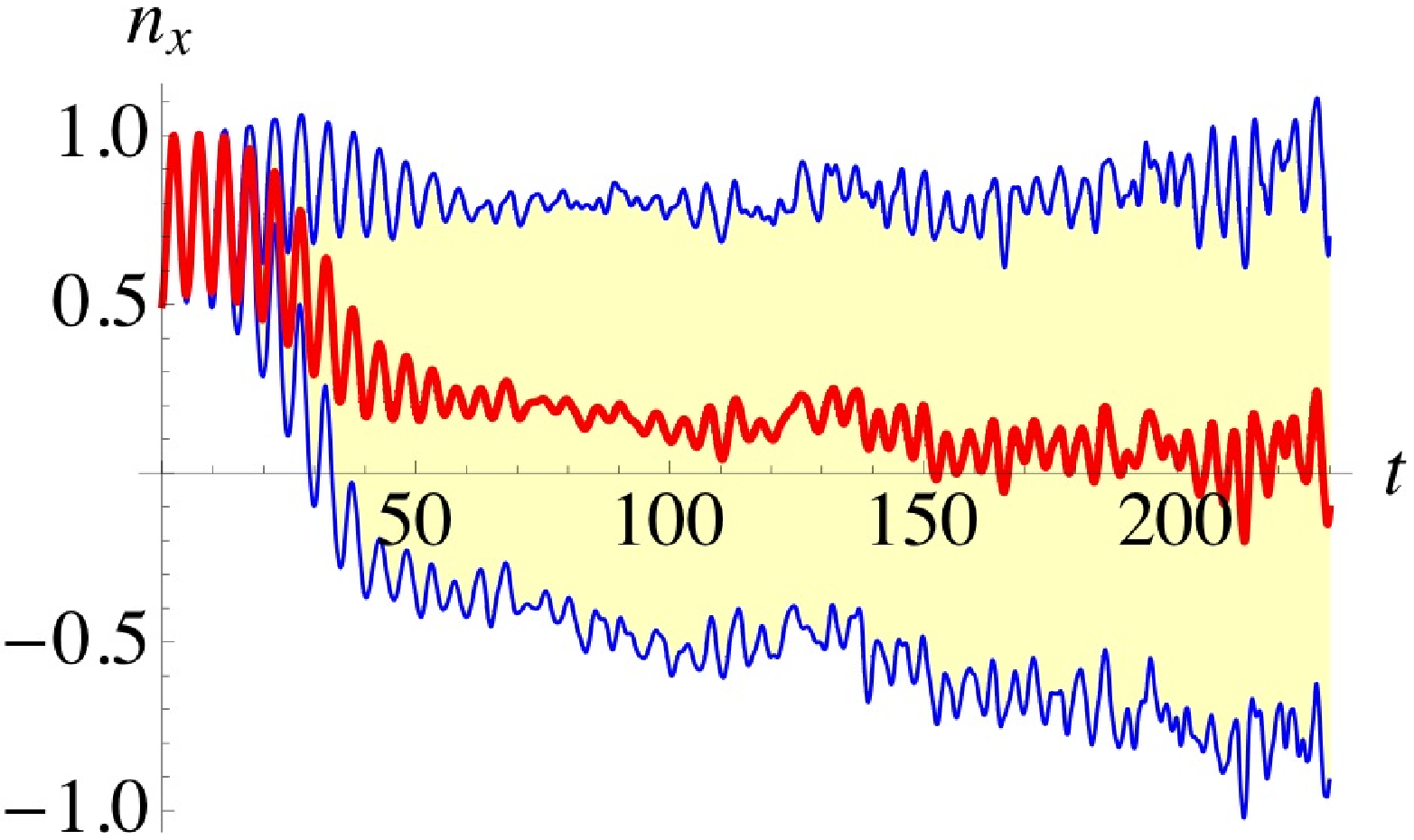}}
\centering\subfigure[]{\includegraphics[width=0.3\textwidth]
{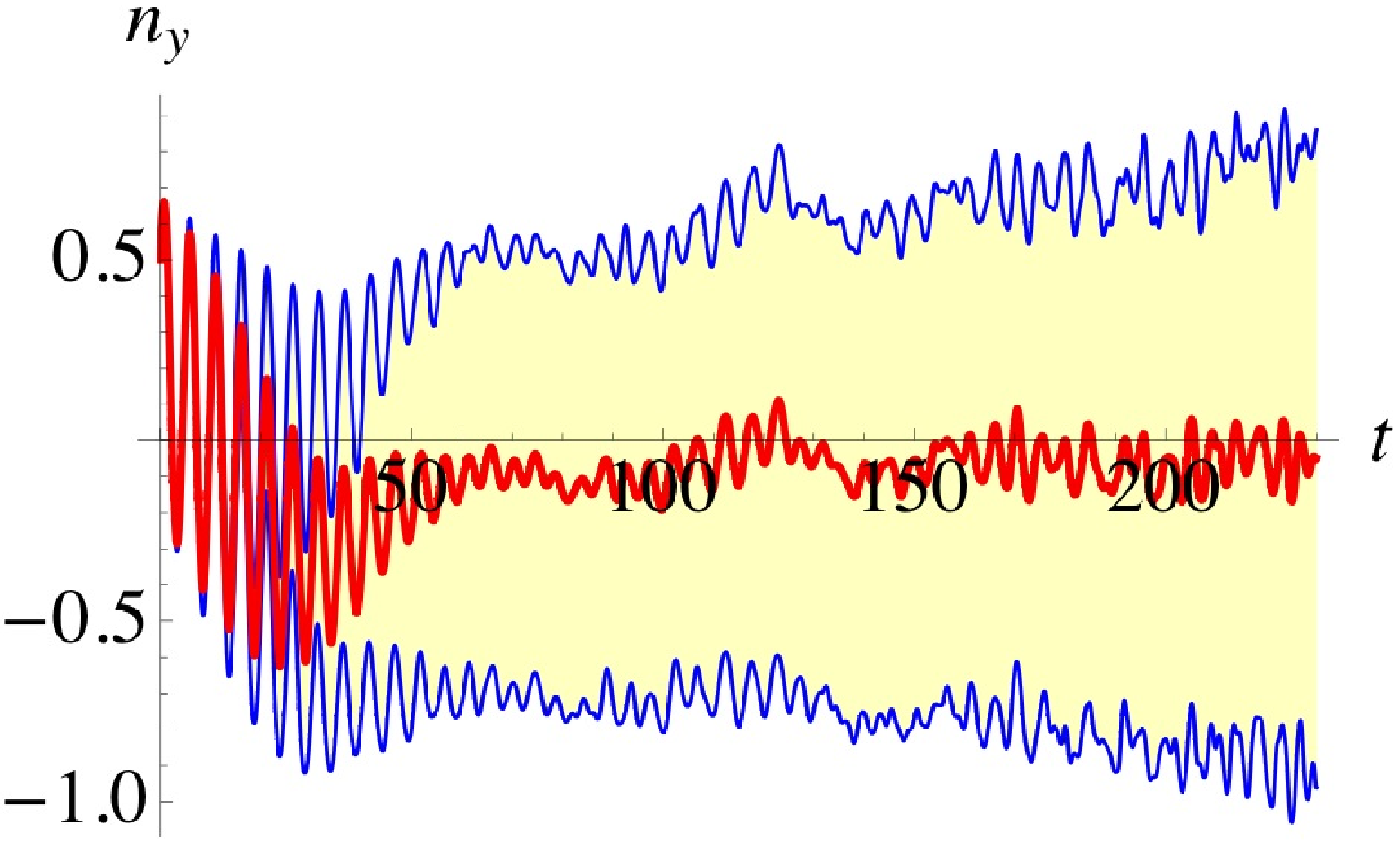}} 
\centering\subfigure[]{\includegraphics[width=0.3\textwidth]
{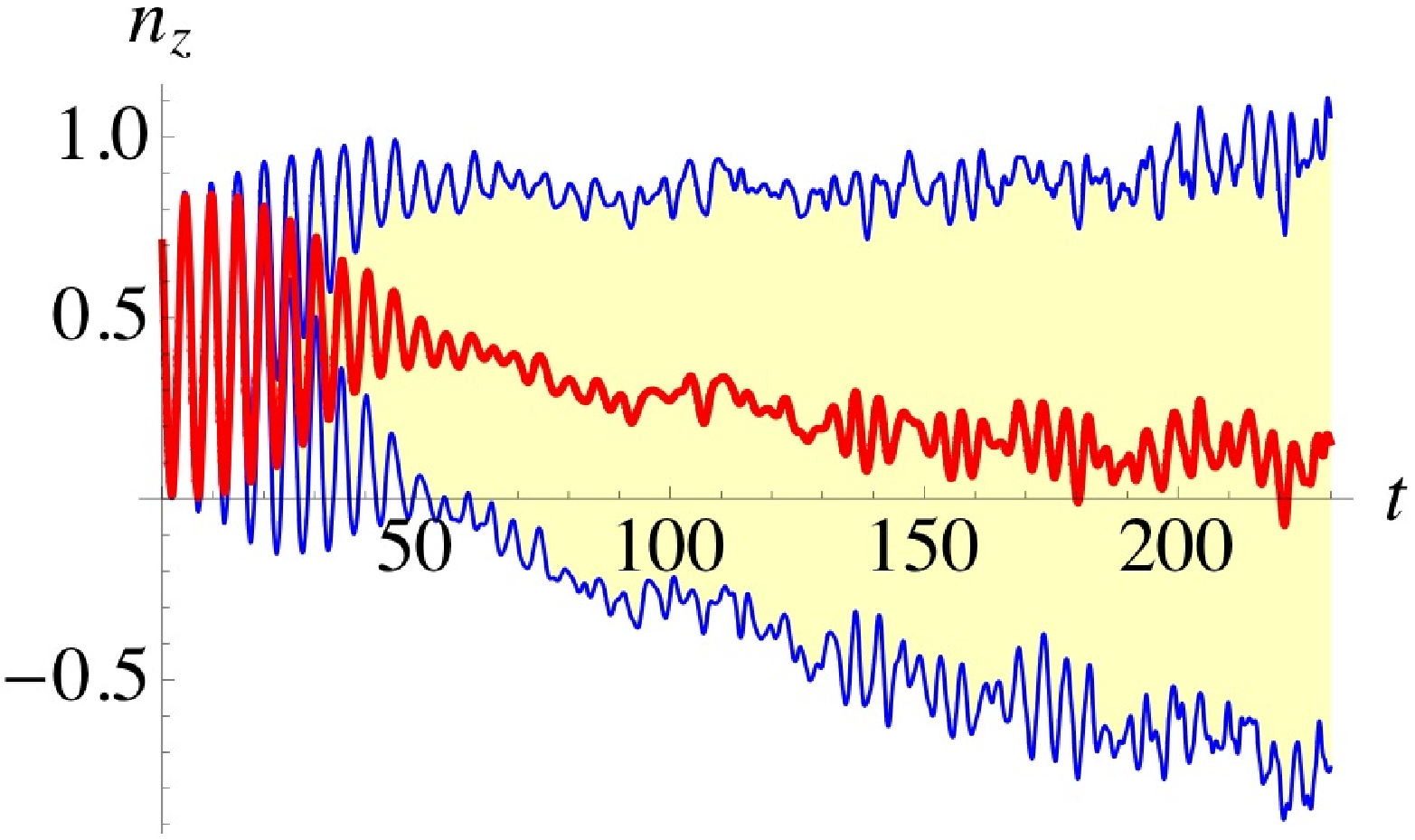}} 
\caption{(Color online) Average and standard deviation of $n_x(t),
n_y(t), n_z(t)$ versus time obtained for stochastic dynamics using
Eqs.~(\ref{dot_dn_final_nl}), (\ref{dot_dL_nl}) and
(\ref{d_varepsilon}), with $\varepsilon_x(t)$, $\varepsilon_y(t)$ and
$\varepsilon_z(t)$ fields taken as Gaussian white noise and $\sigma =
0.1$.}
\label{Fig_induced_n_nonlinear}
\end{figure}

\begin{figure} % [hbt]
\centering
\centering\subfigure[]{\includegraphics[width=0.3\textwidth]
{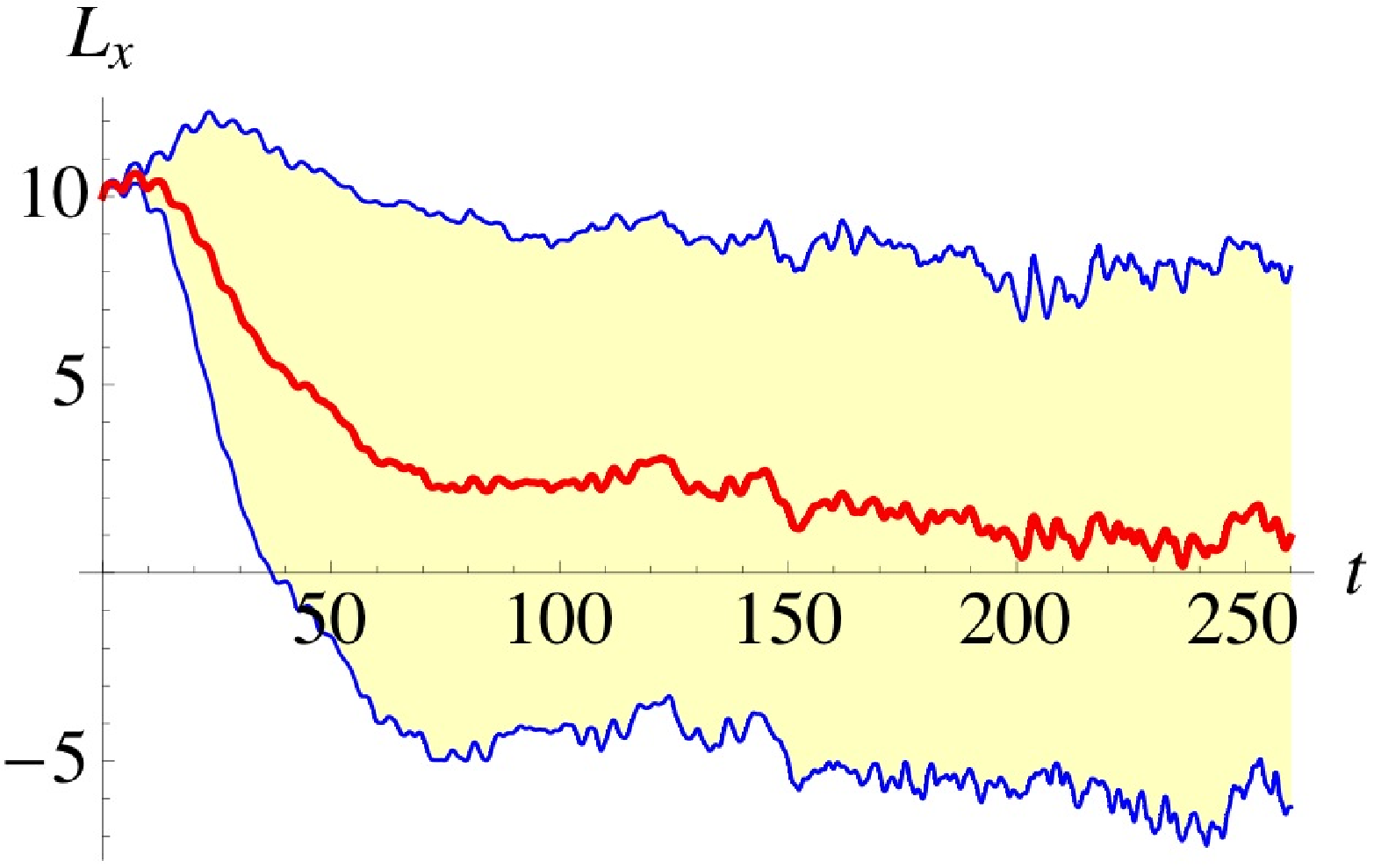}}
\centering\subfigure[]{\includegraphics[width=0.3\textwidth]
{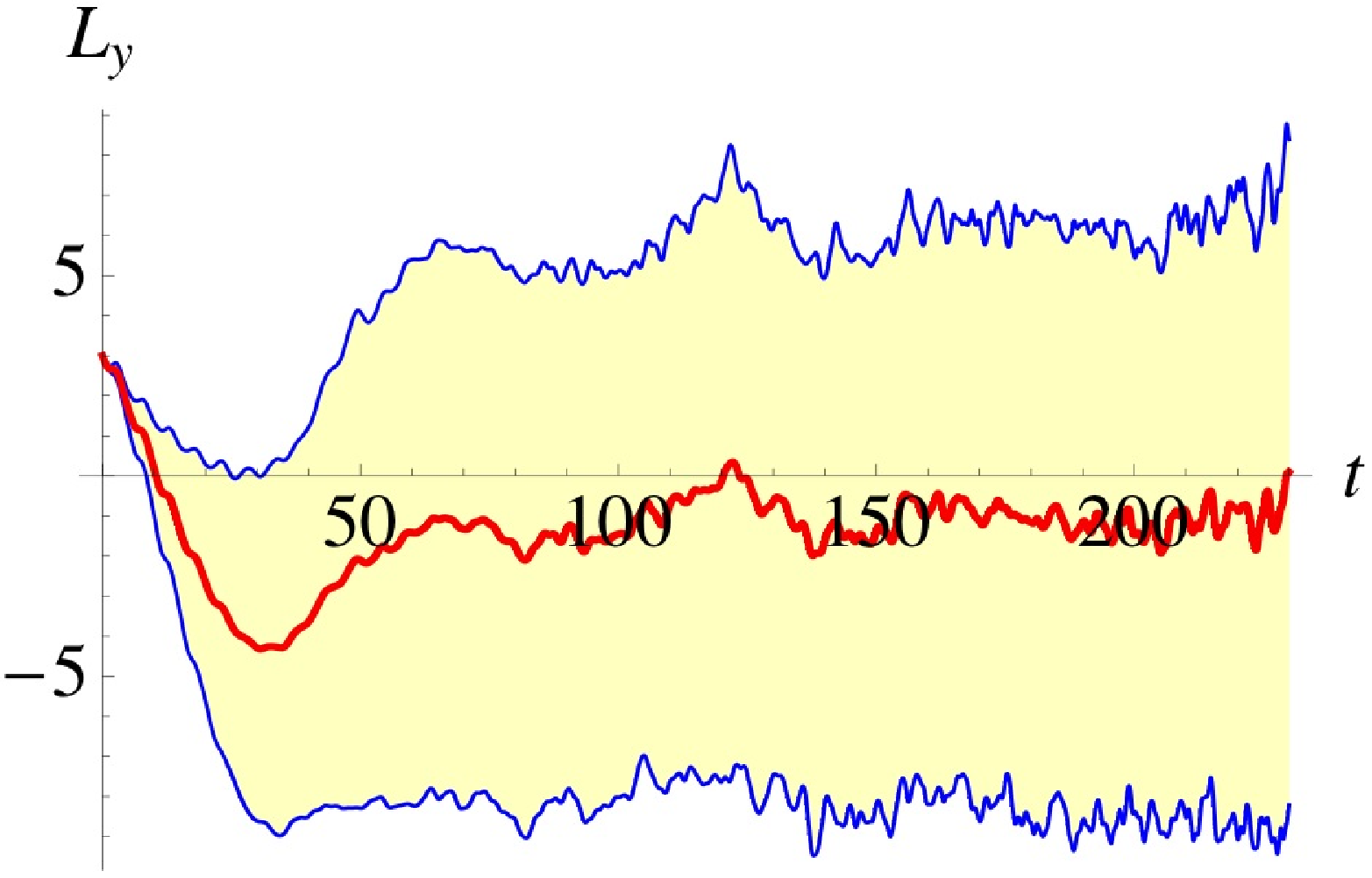}} 
\centering\subfigure[]{\includegraphics[width=0.3\textwidth]
{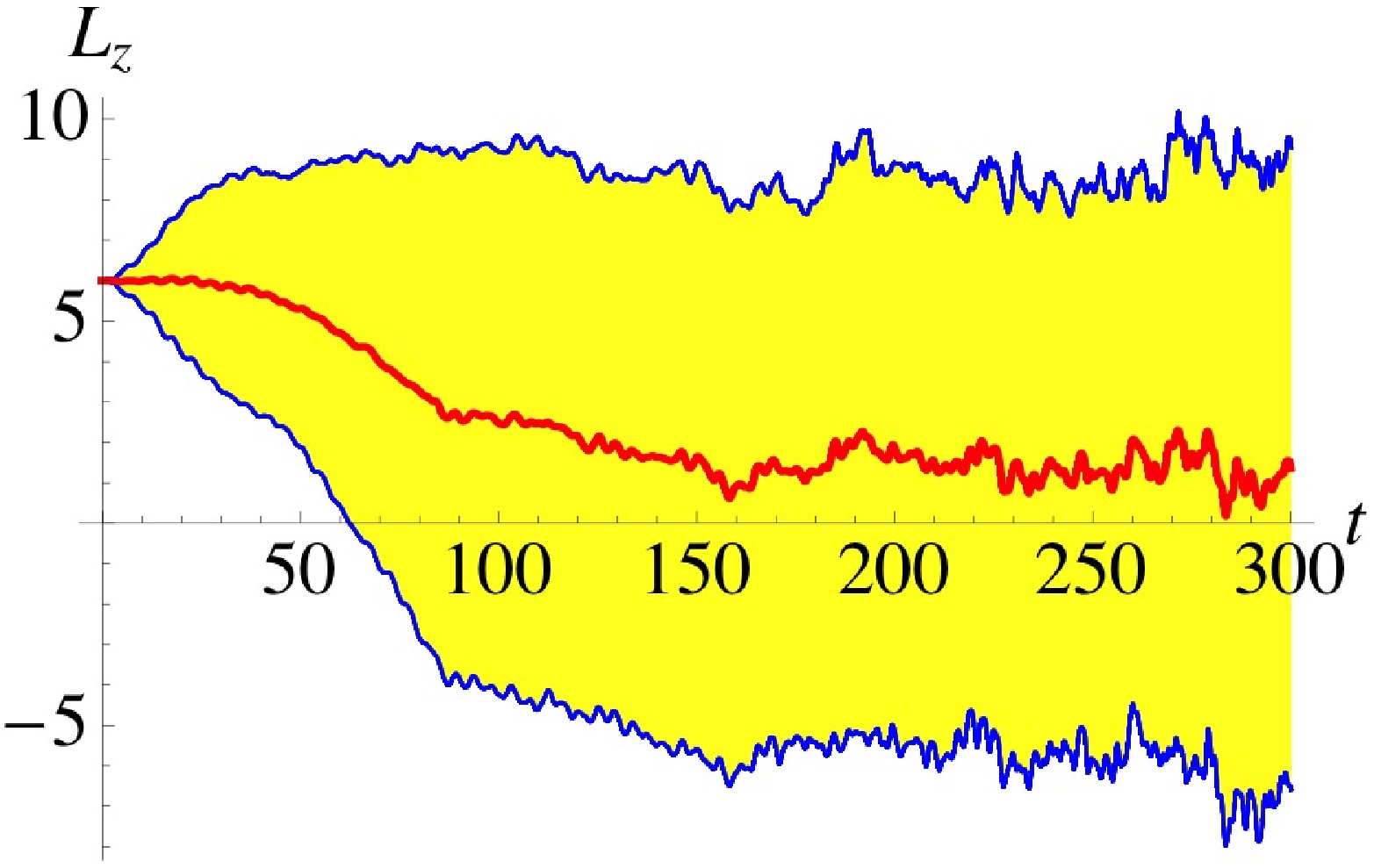}} 

\caption{(Color online) Average and standard deviation of the angular
momentum vector $(L_x(t), L_y(t), L_z(t))$ versus time obtained for
stochastic dynamics using Eqs.~(\ref{dot_dn_final_nl}),
(\ref{dot_dL_nl}) and (\ref{d_varepsilon}), with $\varepsilon_x(t)$,
$\varepsilon_y(t)$ and $\varepsilon_z(t)$ fields taken as Gaussian
white noise and $\sigma = 0.1$.}
\label{Fig_induced_L_nonlinear}
\end{figure}

Figures \ref{Fig_induced_n_nonlinear} and
\ref{Fig_induced_L_nonlinear} show the results of the nonlinear
calculation with a value of $\sigma = 0.1$.  There is a significant
difference between the perturbative results shown in
Figs.~\ref{Fig_vareps_z_n_m_v} and \ref{Fig_vareps_z_L_m_v} and the
nonperturbative results, even for times as early as $t = 25$, and
decoherence is already significant for times beyond $t = 25$.  The
most significant difference is that the dynamical variables $n_z(t)$
and $L_z(t)$ no longer `hang up' at finite values, but decay to zero
at large time.

\section{Summary and Conclusion}   \label{Sec:Summary}

We introduced a model for treating the dynamics of a molecule with an
induced dipole moment in the presence of an external electric field.
We showed that the classical dynamics is equivalent to the mean-field
quantum dynamics of the system.  The dynamics is more complicated than
the dynamics of a magnetic dipole moment in a magnetic field; modeling
the dynamics requires equations of motion for {\em both} the angular
momentum operator $\hat{{\bf L}}(t)$ and the operator for the unit
vector in the direction of the axis of the molecule, $\hat{{\bf
n}}(t)$.  Then, we considered the dynamics in the presence of an
external electric field that is a sum of a deterministic field and a
stochastically fluctuating field (noise).  For simplicity, we took the
fluctuations to be Gaussian white noise.  The model makes the external
noise assumption \cite{vanKampenBook} wherein no back-action of the
system on the environment is present.  Using perturbation theory for
the stochastic field, the $z$ component of the average induced
electric dipole moment, $\langle {\hat{{\bf n}}}(t) \rangle$, and the
$z$ component of the average angular momentum, $\langle {\hat{{\bf
L}}}(t) \rangle$, do not decay (decohere) to zero, despite
fluctuations in all three components of the electric field, (but the
other components of these vectors do decohere).  This is in contrast
to the decay of the average over fluctuations of a magnetic moment,
which does decohere to zero in a stochastic magnetic field with
Gaussian white noise in all three components \cite{STB_2013}.  In
contradistinction to the perturbative analysis, i.e., upon including
the term nonlinear in the stochastic field in the equations of motion,
we find that decoherence occurs in all three components of $\langle
\hat{{\bf L}}(t)\rangle$ and $\langle \hat{{\bf n}}(t)\rangle$.
Moreover, decoherence of the transverse components of these vectors
appears significantly earlier than in the perturbation theory
solutions.  These predictions, obtained under the external noise
assumption, should be able to be readily checked experimentally.
These predictions should remain valid also for Gaussian colored noise
stochastic processes, as long as the temporal correlation time of the
colored noise process, $\tau_c$, is short compared with the rotation
time of the molecule, $\tau_r = I/\langle L \rangle$, and the Stark
timescale, $\tau_S = \hbar/(Ed)$.

\begin{acknowledgments}
This work was supported in part by grants from the Israel Science
Foundation (No.~2011295) and the James Franck German-Israel Binational
Program.  We are grateful to John Scales for useful conversations.
\end{acknowledgments}

\end{document}